\documentclass[pre,twocolumn,showpacs]{revtex4}
\usepackage{graphicx}
\usepackage{dcolumn}
\usepackage{color}
\usepackage{mathrsfs}
\usepackage{amssymb,amsmath}
\usepackage{bm}
\usepackage{epsfig}

\def\beq{\begin{equation}}
\def\eeq{\end{equation}}
\def\bea{\begin{eqnarray}}
\def\eea{\end{eqnarray}}
\begin{document}
\title{Shear Unzipping of DNA}
\author{Buddhapriya Chakrabarti$^{1}$, and David R. Nelson$^{2}$}
\affiliation{${1}$ Department of Polymer Science and Engineering, University of
Massachusetts, Amherst, MA 01003. \\
${2}$ Lyman Laboratory of Physics, Harvard University, Cambridge, MA
02138.}
\email{buddho@polysci.umass.edu}
\email{nelson@physics.harvard.edu}
\begin{abstract}
We study theoretically the mechanical failure of a simple model of double stranded DNA under an applied shear. Starting from a more microscopic Hamiltonian that describes a sheared DNA, we arrive at a nonlinear generalization of a ladder model of shear unzipping proposed earlier by deGennes\cite{deGennes:01}. Using this model and a combination of analytical and numerical methods, we study the DNA ``unzipping'' transition when the shearing force exceeds a critical threshold at zero temperature. We also explore the effects of sequence heterogeneity and finite temperature and discuss possible applications to determine the strength of colloidal nanoparticle assemblies functionalized by DNA.
\end{abstract}
\pacs{}
\date{\today}
\maketitle

\section{Introduction}
\label{intro}
\begin{figure}
\includegraphics[width=8cm,height=7cm]{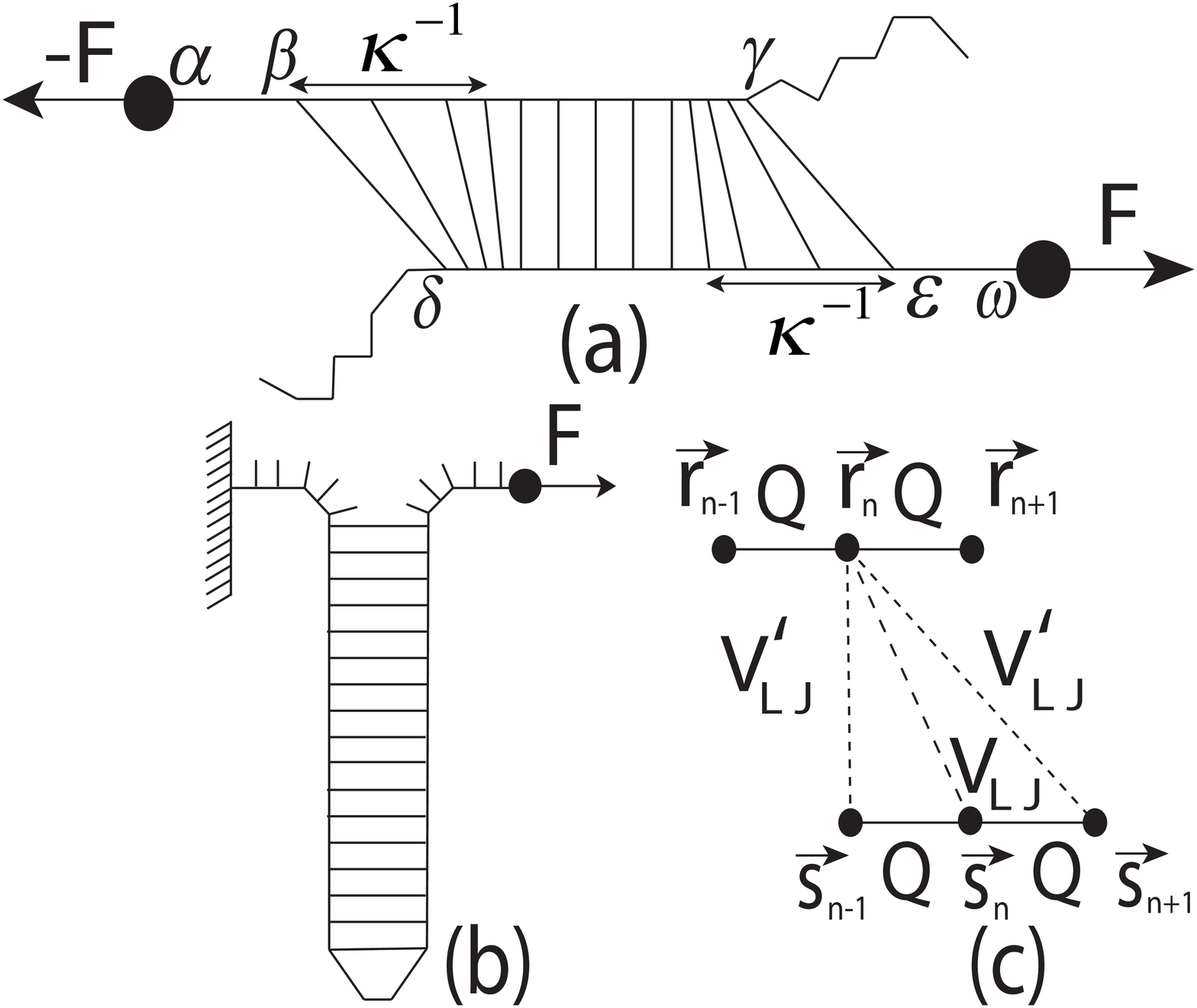}
\caption{\label{Schematic} Schematic figure showing DNA unzipping in a shear mode (a) and in a more conventional tensile mode (b). In the shear mode, $\vec{u}_{n}$ and $\vec{v}_{n}$ (see text) are the displacements defined along the upper $\alpha\beta\gamma$ and lower $\delta\epsilon\omega$ strands respectively. $\vec{F}$ represents the shear force. Panel (c) shows an enlarged section of the sheared configuration. The position co-ordinates of the $n$-th nucleotide on the top and bottom rungs of the ladder are represented by $\vec{r}_{n}$ and $\vec{s}_{n}$ respectively. $V_{LJ}$ and $V^{\prime}_{LJ}$ represent the complementary base pairing and stacking interactions (see text).}
\end{figure}

The physics of single molecule DNA unzipping experiments\cite{Bustamante:03} (in vitro mechanical analogues of helicase-mediated unzipping during DNA replication\cite{Alberts:94}) is fairly well understood. However, much less is known about shear denaturation of DNA (see Fig.\ref{Schematic}). Although of less immediate biological relevance, there are interesting materials science applications such as determining the strength of DNA/gold nanoparticle assemblies. In one implementation\cite{Mirkin:96}, gold nanoparticles are functionalized by attaching single-stranded thiol-capped oligonucleotides. Nanoparticle aggregates form (and the solutions change color) when duplex DNA with complementary ``sticky ends'' is added to the solution. In addition to DNA detection applications associated with a shift in the plasmon absorption frequency upon aggregation\cite{Park:03}, (which depends upon particle size, concentration and interparticle distance) recent attention has focused on thermal denaturation\cite{Kiang:03} and the effect of disorder\cite{Harris:05} when single stranded linker elements are used. In this paper we adapt methods developed for DNA unzipping (generalizing a simple harmonic model of deGennes\cite{deGennes:01}), in an effort to understand the inherent strengths and shear denaturation pathways of the hybrid DNA ``bonds'' that hold these aggregates together at low temperatures.

The physics governing the failure of a double stranded DNA in shear mode is rather different from that in a conventional tensile mode. Consider the two geometries shown schematically in Fig.\ref{Schematic}(a) and \ref{Schematic}(b). In the tensile mode of Fig.\ref{Schematic}(b), bases are sequentially stretched as the duplex is unzipped by forces exerted on the same end of the duplex. For shear unzipping, however, these forces act on opposite ends (as well as on opposite duplex strands), and the stretching is spread out over many base pairs. The statistical mechanics of DNA unzipping in the tensile mode can be simply understood by associating an energy function that depends on the number of unzipped bases $m$, $\varepsilon(m) = m f$ where $f = 2 g(F) - g_0$, $g_0$ being the strength of the bonds and $g(F)$ the elastic response of the unzipped handles including entropy effects\cite{Lubensky:00, Lubensky:02, Nelson:04}. Near the unzipping transition $\textit{\textbf{f}}(F) \sim F - F_c$, where $F$ is the applied tensile force, and $F_c$ the unzipping force of the DNA. As $F \rightarrow F^{+}_{c}$, this analysis leads to the scaling relation $<m> \sim 1/(F - F_c)$ for a homopolymer and $\overline{<m>} \sim 1/(F - F_c)^2$ for heteropolymers\cite{Lubensky:00, Lubensky:02}.

In this paper we explore the shear unzipping transition in the same spirit. As pointed out in Ref.\cite{deGennes:01}, an important length scale for the DNA is the length $\kappa^{-1}$ over which strain relaxes on either end (see Fig.\ref{Schematic}(a)). Experiments that measure rupture forces of short hybridized DNA segments with one strand grafted onto a substrate at one end, and the other strand grafted to a tip of an AFM at the other end show an increasing dependence of the rupture force on the overlap length of the two strands\cite{Lee:94,Mazzola:99,Csaki:01}. In addition, the unzipping force for a 20 base pair long sequence is about three orders of magnitude larger than that of the rupture force of a single bond\cite{Lee:94}. Using a simple model of harmonic springs, deGennes showed that the rupture force scales linearly with the overlap length for small system sizes while it saturates for large ones\cite{deGennes:01}. Recently, these predictions have been verified experimentally by the Prentiss group\cite{Danilowicz:08,Hatch:08}. In these experiments one end of a DNA hybrid is grafted to a glass capillary by antigen-antibody linkages while the complimentary strand at the opposite end is attached to a magnetic bead via biotin-streptavidin linkage. The magnetic bead is then placed in a uniform magnetic field gradient to ensure that a constant shear force is applied to the DNA. Here, we elaborate the deGennes model, and study whether the physics of shear unzipping coarse-grained over this strain relaxation length yields the same physics as conventional unzipping in context of a simple model Hamiltonian. We answer this question in the affirmative. With the help of a ``semi-microscopic'' model of a poly-(AT) hybrid, we find that the strain is indeed localized over a narrow region $\kappa^{-1}$ on either sides of the chain. At low temperatures the chain unzips when the force exceeds a threshold value $F \approx f_0 L$ for short chains and $F \approx 2 f_0/\kappa a$ for long ones, $f_0$ being the rupture force of a single bond, $L$ the system size (measured in units of the average nucleotide) and $a$ the equilibrium spacing between the bases along the backbone. This behavior is similar to that observed in homopolymer unzipping in a tensile mode, albeit with a different critical unzipping force. We find that the simplified ladder model of deGennes can be obtained from a more general nonlinear Hamiltonian by modeling the sugar phosphate backbone by Hookean springs and expanding in the displacements of (i) the nearest neighbor interactions among complementary base pairs on opposite rungs of the ladder and (ii) the next nearest neighbor interactions on different rungs of the ladder (a simple model of stacking interactions) to quadratic order and projecting out the eigenmode corresponding to shear deformation. We then study shear unzipping of a heteropolymer taking cues from previous studies done in tensile mode. We characterize the failure pathway of the heteropolymer in terms of a meandering exponent that measures how different it is from that of a homopolymer where the outermost bonds on the left and right half of the chain break in alternation. Finally, we consider a nonlinear vector model of a poly-(AT) hybrid. The results from this analysis are qualitatively similar to that of the simplified model proposed in Ref.\cite{deGennes:01}, although the mode of failure requires the breaking of cross-braces, which are required to stabilize the vector model.
The cross-braces eliminate a spurious soft mode present in the most straightforward generalization of the deGennes model\cite{footnote}. Our most important conclusion is that the physics of a vector model of shear unzipping of homopolymers at zero temperature reduces to a nonlinear generalization of the original deGennes model at long wavelengths.

\section{Semi-microscopic vector model}
\label{semimicroscopicvectormodel}
To introduce the basic ideas, consider a simple nonlinear model applicable to the poly-(A:T)-hybrid shown in Fig.\ref{Schematic}. If $\vec{r}_{n}$ and $\vec{s}_{n}$ are the position vectors of the $n$-th nucleotide along the upper and lower strands respectively, and $L$ is the size of the overlap region (measured in units of the average nucleotide-spacing), the energy of a sheared configuration is
\begin{eqnarray}\label{Semi-microscopic-H-Eq}
H=\frac{1}{2} Q \sum^{L}_{n=1} \left(
\left[|\vec{r}_{n+1}-\vec{r}_{n}|- a\right]^{2}+
\left[ |\vec{s}_{n+1}-\vec{s}_{n}|-a \right]^{2} \right) + \nonumber \\
\sum^{L}_{n=1}\left[ V_{LJ}(|\vec{r}_{n} - \vec{s}_{n}|) +
V^{\prime}_{LJ}(|\vec{r}_{n+1} - \vec{s}_{n}|) +
V^{\prime}_{LJ}(|\vec{s}_{n+1} - \vec{r}_{n}|) \right],
\end{eqnarray}
where $Q$ (see Fig.\ref{Schematic}) is a spring constant representing the stiff sugar-phosphate backbones. When $Q$ is large, one can safely neglect the elastic properties of the single stranded spacer elements $\alpha\beta$ and $\epsilon\omega$. The direct A:T nucleotide pairing is described for simplicity by a nonlinear Lennard-Jones pair potential
\begin{eqnarray}\label{Lennard-Jones-Potential-Eq}
V_{LJ}(|\vec{r}_{n}-\vec{s}_{n}|) = 4 \epsilon \left[
\left(\frac{\sigma}{|\vec{r}_{n}-\vec{s}_{n}|} \right)^{12}-
\left(\frac{\sigma}{|\vec{r}_{n}-\vec{s}_{n}|} \right)^{6} \right].
\end{eqnarray}
The cross-brace interactions, $V^\prime_{LJ}(|\vec{r}_{n+1} - \vec{s_{n}}|)$ and $V^\prime_{LJ}(|\vec{s}_{n+1} - \vec{r_{n}}|)$, represent the inter-strand next-nearest neighbor interactions among nucleotides (shown as dashed lines in Fig.\ref{Schematic}). We take these interactions to be described by a Lennard-Jones form as well, but with a different well depth and spatial extent. Thus
\begin{eqnarray}\label{Cross-Braces-Potential-Eq}
V^{\prime}_{LJ}(|\vec{r}_{n+1}-\vec{s}_{n}|)= 4 \epsilon^\prime \left[ \left(\frac{\sigma^\prime}{|\vec{r}_{n}-\vec{s}_{n}|} \right)^{12} - \:\: \left(\frac{\sigma^\prime}{|\vec{r}_{n}-\vec{s}_{n}|} \right)^{6} \right],
\end{eqnarray}
with well depth $\epsilon^\prime \approx 1.6 \,k_B T$,\cite{Sponer:97} and a length scale $\sigma^\prime = 2^{-1/6} \sqrt{\tilde{\sigma}^2 + a^2}$, where $\tilde{\sigma} = 2^{1/6} \sigma$ the equilibrium separation between the two strands. These cross braces play the role of stacking energies, and stiffen the DNA ladder against an applied shear and remove a soft mode such that the inter-strand separation decreases till the ladder collapses\cite{footnote}. Since the stacking interactions represented by our cross braces involve base pairs ``touching'' over a larger surface area, we assume that these interactions are somewhat stronger than the complimentary interactions. Adding opposing forces $\vec{F}$ to the two opposing strands at opposite ends (Fig.\ref{Schematic}(a)) leads to total energy $H^\prime = H - \vec{F} \cdot (\vec{r}_{1} - \vec{s}_{L})$. Upon defining $\vec{u}_{n}$ and $\vec{v}_{n}$ as the displacement fields on the top and bottom rungs of the ladder, the position co-ordinates $\vec{r}_{n}$ and $\vec{s}_{n}$ can be written in cartesian coordinates as $\vec{r}_{n} = n a \hat{x} + (1/2) \tilde{\sigma} \hat{y} + \vec{u}_{n}$ and $\vec{s}_{n} = n a \hat{x} - (1/2) \tilde{\sigma} \hat{y} + \vec{v}_{n}$ respectively, $a$ being the equilibrium unstretched separation between nucleotides on the two sugar phosphate backbones. When thermal fluctuations are neglected, the equilibrium physics of such a ladder is obtained by solving the over-damped equations of motion for the
phonon fields $\vec{u}_{n}$,
\begin{equation}\label{Equations-Of-Motion-Eq}
\frac{\partial \vec{u}_{n}}{\partial t} = - \Gamma \frac{\delta
H^\prime}{\delta \vec{u}_{n}},
\end{equation}
and similarly for $\vec{v}_{n}$, stepping forward in time until a steady state is reached. In Eq.(\ref{Equations-Of-Motion-Eq}), $\Gamma$ sets the inverse time scale over which the system equilibrates.

An understanding of the equilibrium properties {\it of two} DNA hybrids like that in Fig.\ref{Schematic} connected in series would be necessary to describe the aggregates of Refs.\cite{Mirkin:96,Park:03,Kiang:03,Harris:05}. Although more microscopic models are certainly possible\cite{Lavery:99}, the Hamiltonian $H^\prime$ is simple enough to allow both analytic insights and an evaluation of the effect of sequence heterogeneity, modeled by nonlinear pair potentials with \textit{sequence specific} Lennard-Jones binding energies $\epsilon_{n}$ and $\epsilon^{\prime}_{n}$. While the model neglects helical twist, Lavery and Lebrun\cite{Lavery:99} have shown from an \textit{ab initio} approach that forces acting on opposing $3^\prime$ ends of a double stranded DNA segment eventually lead to a planar configuration, much like that assumed in Ref.\cite{deGennes:01} and sketched in Fig.\ref{Schematic}(a).

We postpone a full nonlinear analysis of sheared DNA described by the semi-microscopic Hamiltonian described in Eq.(\ref{Semi-microscopic-H-Eq}) to Sec.\ref{nonlinearscalardisplacementmodel}. Instead, we first expand the Hamiltonian up to quadratic order in the displacement fields $\vec{u}_{n}$ and $\vec{v}_{n}$. We then compute the eigenfrequencies and identify the low energy mode corresponding to the shear deformation. Up to an additive constant, the Hamiltonian of Eq.(\ref{Semi-microscopic-H-Eq}) for small displacements reads
\begin{eqnarray}
&H&=\frac{1}{2} Q \sum^{L}_{n=1} (u^{x}_{n+1} - u^{x}_{n})^2 + \frac{1}{2} Q \sum^{L}_{n=1} (v^{x}_{n+1} - v^{x}_{n})^{2} \nonumber \\ &+& \frac{36 \epsilon}{\tilde{\sigma}^{2}} \sum^{L}_{n=1} (u^{y}_{n} - v^{y}_{n})^{2} + 18 \epsilon^\prime \sum^{L}_{n=1} \left[\frac{a^2 (u^{x}_{n+1} - v^{x}_{n})^{2}}{(a^2 + \tilde{\sigma}^2)^2} \right. \nonumber \\ &+& \left. \frac{\tilde{\sigma}^2 (u^{x}_{n+1} - v^{x}_{n})^{2}}{(a^2 + \tilde{\sigma}^2)^2} + \frac{2 a \tilde{\sigma} (u^{x}_{n+1} - v^{x}_{n})(u^{y}_{n+1} - v^{y}_{n})}{(a^2 + \tilde{\sigma}^2)^2} \right] \nonumber \\ &+& 18 \epsilon^{\prime} \sum^{L}_{n=1} \left[ \frac{a^2 (v^{x}_{n+1} - u^{x}_{n})^{2}}{(a^2 + \tilde{\sigma}^2)^2} + \frac{\tilde{\sigma}^2 (v^{x}_{n+1} - u^{x}_{n})^{2}}{(a^2 + \tilde{\sigma}^2)^2} \right. \nonumber \\ &+& \left. \frac{ 2 a \tilde{\sigma} (v^{x}_{n+1} - u^{x}_{n})(v^{y}_{n+1} - u^{y}_{n})}{(a^2 + \tilde{\sigma}^2)^2} \right].
\label{Semi-Microspic-H-Quad-Approx-Eq}
\end{eqnarray}
To identify the eigenmode corresponding to shear deformation we numerically diagonalize Eq.(\ref{Semi-Microspic-H-Quad-Approx-Eq}) with free boundary conditions on all the nodes. We then solve the over-damped equations of motion Eq.(\ref{Equations-Of-Motion-Eq}) corresponding to the Hamiltonian of Eq.(\ref{Semi-Microspic-H-Quad-Approx-Eq}) with a shearing term $ - \vec{F} \cdot (\vec{r}_{1} - \vec{s}_{L})$ and match the motion with the modes obtained from the diagonalization procedure.

Fig.\ref{deGennesPhonons} shows the acoustic and optical phonon branches of the corresponding eigenmodes of the Hamiltonian in Eq.(\ref{Semi-Microspic-H-Quad-Approx-Eq}) with the parameters of B-DNA ($\epsilon \approx 1 k_{B} T$, $\tilde{\sigma} = 2.073 A^\circ$). The strength of the stacking interactions $\epsilon^\prime$ was taken as $1.6$ $k_{B} T$. The longitudinal and transverse branches are shown in Fig.\ref{deGennesPhonons}(a) while the optical branches are shown in Fig.\ref{deGennesPhonons}(b). Also shown are the schematic pictures of the configurations of the ladder eigenmodes corresponding to the respective phonon branches. The phonon branch that is flat near $k=0$ is clearly the one excited most prominently by a force-induced shear deformation like that in Fig.\ref{Schematic}. The relative displacement field $\delta^{x}_{n} \equiv u^{x}_{n} - v^{x}_{n}$ along the direction of shear (shown in Fig.\ref{deGennesPhonons}(c)) is visually similar to the one obtained from the elastic description of the simplified model studied by deGennes\cite{deGennes:01}. Fig.\ref{deGennesPhonons}(d) shows the displacement field in the direction perpendicular to the chain axis $\delta^{y}_{n} = u^{y}_{n} - v^{y}_{n}$. Though the displacements are an order of magnitude smaller than those in the direction of shear, we see a reverse deGennes effect: The strain field $\delta^{y}_{n}$ is negative near the edges and zero in the interior implying a vertical distance smaller than the equilibrium spacing $\tilde{\sigma}$ near the edges than in the middle except for the penultimate bond on either side. This pinching effect is a result of the imbalance in the number of next nearest neighbor interactions on the last and last-but-one bond. Having identified the eigenmode corresponding to the shear deformations, we now focus on a nonlinear scalar Hamiltonian that describes this mode alone.

\section{Nonlinear scalar displacement model}
\label{nonlinearscalardisplacementmodel}
A nonlinear generalization of the deGennes Hamiltonian of Ref.\cite{deGennes:01}, written in terms of scalar local phonon fields $u_{n}$ and $v_{n}$ (corresponding to vector phonon displacements projected along the chain direction) is given by
\begin{eqnarray}
H&=&\frac{1}{2} Q \sum^{L}_{n=1} (u_{n+1} - u_{n})^2 + \frac{1}{2} Q
\sum^{L}_{n=1} (v_{n+1} - v_{n})^2 \nonumber \\ &+& \sum^{L}_{n=1}
V_{LJ} \left( |u_n - v_n| \right) - F u_1 + F
v_L.\label{deGennes-Nonlinear-Hamiltonian-Eq}
\end{eqnarray}
Eq.(\ref{deGennes-Nonlinear-Hamiltonian-Eq}), could, in principle be obtained from the more microscopic model of Eq.(\ref{Semi-microscopic-H-Eq}) by solving and eliminating all phonon displacements except those that couple directly to force. We expect that this procedure would lead directly to Eq.(\ref{deGennes-Nonlinear-Hamiltonian-Eq}) with small renormalizations of the spring constants $Q$ and an effective nonlinear pair potential coupling the two strands. The nature of interaction among complementary base pairs in this model is again assumed to be of the same Lennard-Jones form as in Eq.(\ref{Lennard-Jones-Potential-Eq}) \textit{i.e.} $V_{LJ}(|u_n - v_n|) = 4 \epsilon \left( \left(\sigma/|u_n - v_n| \right)^{12} - \left(\sigma/|u_n - v_n| \right)^{6} \right) $, with renormalized parameters $\epsilon$ and $\sigma$ . The original model of Ref.\cite{deGennes:01} follows from expanding the potential around its minimum and retaining terms up to quadratic order in the displacement fields. In the harmonic approximation, $V_{LJ} \approx \frac{1}{2} R (u_n - v_n)^2$, and the shear strain $\delta_n = u_n - v_n$ due to a force $F$ is given by,\cite{deGennes:01}
\begin{equation}
\delta_n = \delta_0 \cosh[\kappa n],
\label{delta-n-Solution-deGennes-Eq}
\end{equation}
where $\delta_0 = F/\left(Q \cosh[\kappa L/2] + 2 R \sinh[\kappa L/2]\right)$, and the important elastic screening length $\kappa^{-1}$ expressed in units of the nucleotide spacing is given by the ratio of the strengths of the covalent and hydrogen bonds, $\kappa^{-1} = \sqrt{2 R/Q}$ $\approx 10 a$.

\begin{figure}
\includegraphics[width=8cm,height=7cm]{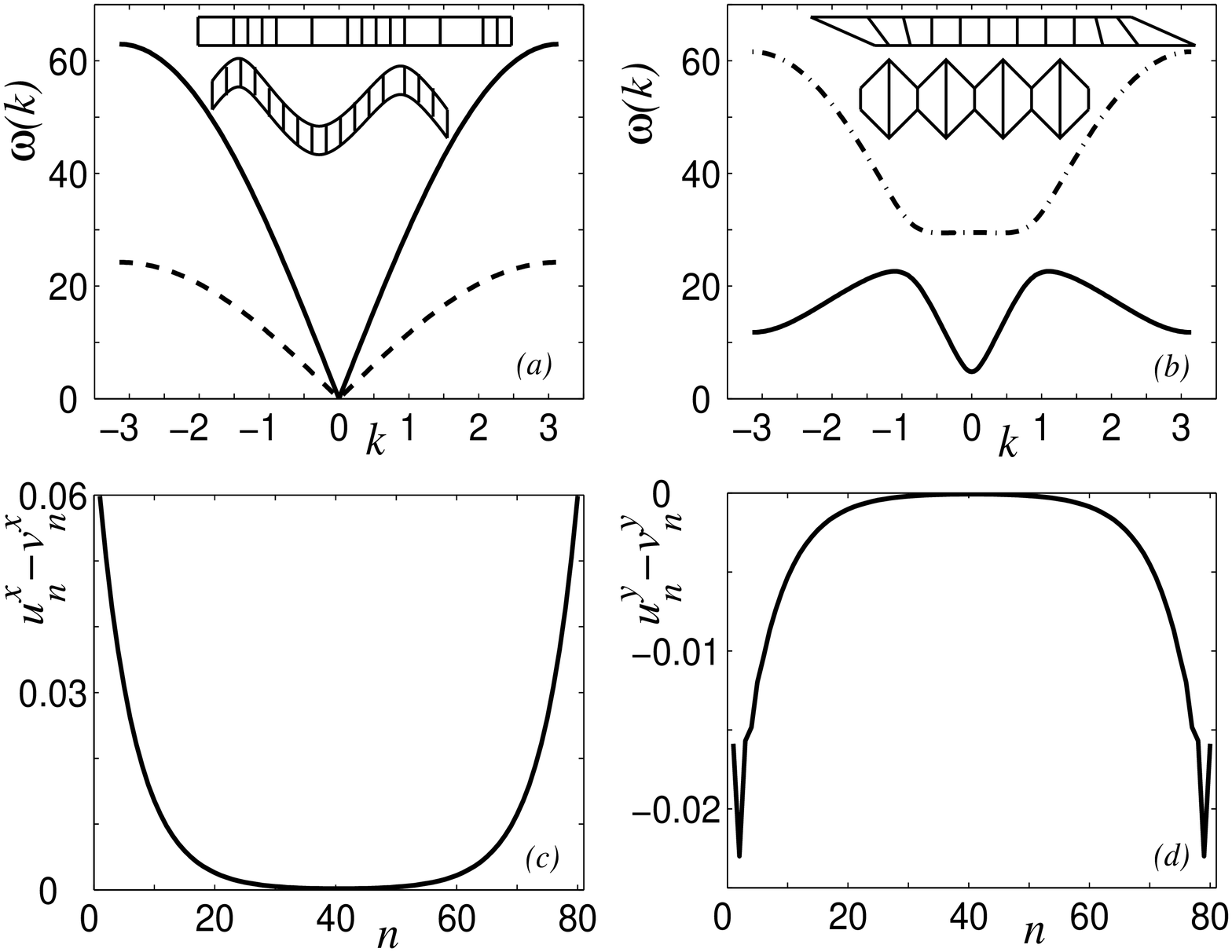}
\caption{\label{deGennesPhonons} Phonon spectrum of a DNA ladder with stacking interactions and the corresponding strain profile. Panel (a) shows the longitudinal (upper) and transverse (lower) acoustic branches while (b) shows the optical branches corresponding to a shear mode (upper curve) and a ``peristalsis'' mode (lower curve). Panel (c) shows the strain profile of the uppermost eigenmode in (b) projected along the direction of the chain. This is the mode studied in Ref.\cite{deGennes:01}. Panel (d) shows the relative displacements in the direction perpendicular to the strain axis. These displacements are much smaller than in panel (c).}
\end{figure}

The strain profile in Fig.\ref{deGennesPhonons}(c), as well as the one obtained by numerical solution of the equations of motion of scalar displacement fields $u_n$ and $v_n$ for the Hamiltonian in Eq.(\ref{deGennes-Nonlinear-Hamiltonian-Eq}), can be fit to Eq.(\ref{delta-n-Solution-deGennes-Eq}). The healing length $\kappa^{-1}$ for the nonlinear generalization of the deGennes model is given by $\kappa^{-1} = \sqrt{2R_{eff}/Q}$ where $R_{eff} = 72 \epsilon/(2^{1/3} \sigma^2)$, the curvature of the Lennard-Jones potential at its minimum. For the semi-microscopic vector model with cross braces a closed form expression for the healing length is more difficult. However, a comparison between the numerical values of the length scale obtained by fitting the strain profile to the form in Eq.(\ref{delta-n-Solution-deGennes-Eq}) with that obtained from fitting the optical phonon branch corresponding to the shear mode at $k=0$ to the form $\omega(k) = A + B k^2$ ($\kappa^{-1} = \sqrt{B/A}$) matches closely.

In the original description of Ref.\cite{deGennes:01}, where complementary base pair interactions were modeled as Hookean springs, the rupture force was not directly calculable, but was instead inserted by hand. The nonlinear generalization proposed here allows us to explore rupture in more detail. Consider first a single nucleotide pair interacting via the Lennard-Jones potential $V_{LJ}$, and acted on by a force $F$. The resulting nucleotide potential is given by $V_{tot} = V_{LJ}(r) - F \cdot r$. When thermal fluctuations are neglected, the bond ruptures as the limit of metastability of the potential $V_{tot}$ is reached. This happens when the slope at the point of inflection matches the force, \textit{i.e.} $V^{\prime \prime}_{tot}(r) = 0$, and
$V^{\prime}_{tot}(r) = F$. For parameters used in this paper, the
rupture force of a single nucleotide pair is $f_0 \approx 5$ pN.
\begin{figure}
\includegraphics[width=8cm,height=7cm]{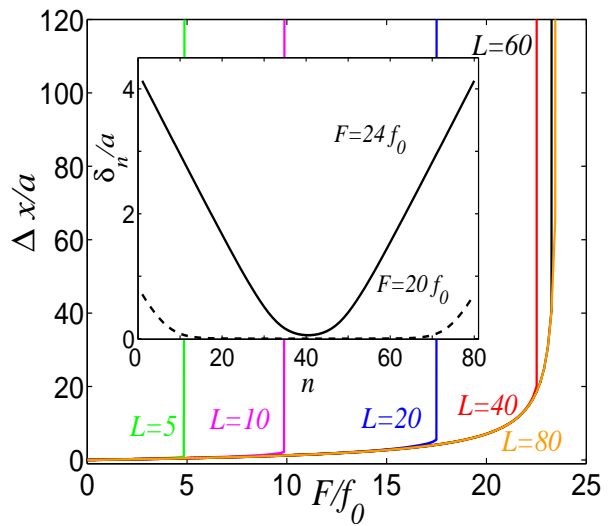}
\caption{\label{figure3} Force (normalized by the rupture force of a single bond, $f_0 \approx 5$ pN) vs.\ extension curves (normalized by the separation of a base pair $a$) for the nonlinear generalization of the deGennes model for different hybridized lengths $L=5$, $10$, $20$, $40$, $60$, $80$. Chain rupture is indicated by the steep upturns in $\Delta x$. Note that the rupture force begins to saturate for large $L$, as predicted in Ref.\cite{deGennes:01}. For the parameters of our model $\kappa^{-1} \approx 10 a$. Inset shows the strain profiles $\delta_n$ for a $L = 80$ sized system for $F = 20 f_0$ (dashed line) and $F = 24 f_0$ (full line).}
\end{figure}
Fig.\ref{figure3} shows the force vs.\ extension curves for DNA hybrids of different overlap lengths. The force is normalized by the unzipping force of a single bond while the total extension $\delta_n = u_n - v_1$ by the equilibrium nucleotide spacing $a$. A system size dependence of the rupture force (denoted by the value at which the extension $\Delta x$ diverges) is observed. Although for small hybridized DNA segments $L << 1/ \kappa a$, the rupture force $F_c = f_0 L$, for large system sizes $L >> 1/\kappa a$ one expects that $F_c \approx 2 f_0/ \kappa a$. This saturation of the rupture forces can be explained as follows\cite{deGennes:01}. For short hybridized DNA segments, when a shear force is applied all the bonds are stretched. Since all the springs are acting in approximately parallel the effective spring constant of the system is given by $k_{eff} = \sum_{i} k_i$, $k_i$ being the spring constant of the $i^{th}$ spring. Hence the force required to break the DNA hybrid is given by $F_c = f_0 L$. For long DNA strands, the strain relaxes over a length $\kappa^{-1}$ on either side. Thus approximately $1/\kappa a$ springs act in parallel on either side and hence the rupture force is $2 f_0 /\kappa a$. A scaling function that interpolates between small and large overlap lengths is given by $F_c = 2 f_0 /\kappa a (1 - 2 \exp[-\kappa L])$. The inset shows the strain profile for a $L=80$ base pair DNA ladder with a shear force $F=20f_0$ and $F=24f_0$. Although these profiles can be fit to Eq.(\ref{delta-n-Solution-deGennes-Eq}) near the center of the construct, the strain profile is in fact \textit{linear} near the edges of the ladder with a slope $F/Q$. The physics in this regime is described by a partially unzipped chain, with some of the force taken up by the stretched sugar phosphate backbone.

\section{Failure pathways for shear unzipping}
\label{failurepathwaysforshearunzipping}
\begin{figure}
\includegraphics[width=8cm,height=7cm]{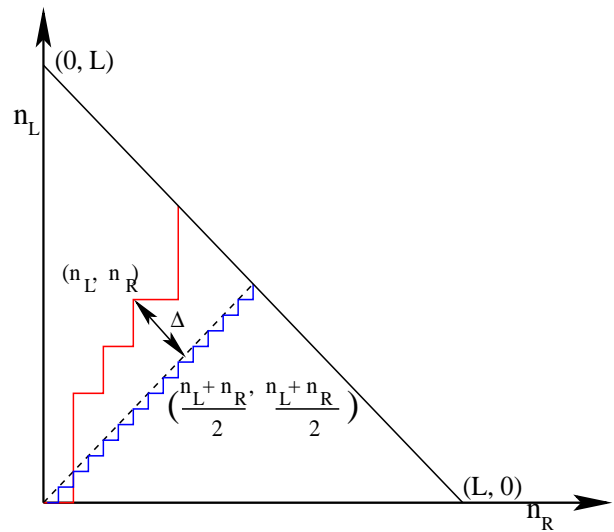}
\caption{\label{figureX} Schematic figure showing the meandering of the failure pathway of a heteropolymer (red line). The meandering is parametrized by the deviation $\Delta$ of the failure pathway from that of a homopolymer (dashed line). The failure pathway for a homopolymer (dashed line) is an average over many runs. On average, we can imagine that the bonds break in alternation and a staircase like pathway (blue line) results either on the upper half or the lower half, depending upon whether the outermost bond is broken on the left or the right side. In contrast to the left-right bond breaking we expect on average in a homopolymer, repeated ruptures of the same heteropolymer should follow the same failure pathway.}
\end{figure}

The strain profile of a homopolymer (see Fig.\ref{deGennesPhonons}(c) and Fig.\ref{figure3}) is symmetric about the
midpoint of the chain as a consequence of the $n \rightarrow - n$ symmetry of the Hamiltonians in Eqs.(\ref{deGennes-Nonlinear-Hamiltonian-Eq}) and (\ref{Semi-microscopic-H-Eq}). However such a symmetry is absent for a heteropolymer. The quenched heterogeneity in the bond strengths is reflected in the failure pathway, a template of the unzipping process in $n_L - n_R$ space, where $n_L$ indicates the number of bonds broken on left half of the chain and $n_R$ the number on the right (see Fig.\ref{figureX}). For a homopolymer, we can summarize the result of averaging over many runs by observing that the left and right bonds break in alternation from the outside edges of the duplex inwards. The failure pathway can be different, however, when modest bond heterogeneity is present. Although we expect that bonds still break from the outside in, several bonds might break first on the right side, followed by a different number on the left \textit{etc}. Provided thermal fluctuations can be neglected, this failure pathway provides a reproducible fingerprint of the rupture process. The failure pathway in the $n_L - n_R$ space resembles that of a $2$-d random walk away from the diagonal line $n_L = n_R$ as the failure progresses, a process controlled by the heterogeneity in the DNA sequence.

As a measure of the randomness of the failure pathway we compute the ``meandering exponent'' that measures how much the failure pathway deviates from that of a homopolymer as a function of the total number of steps. A schematic representation of the failure pathway is shown in Fig.\ref{figureX}. The dashed line passing through the origin denotes the average failure pathway of a homopolymer (we take this to be an average of the ``staircase'' pathway shown in blue either on the upper half plane or the lower half plane depending upon whether the left most or the right most outer bond is broken first). The x-axis and y-axis denote the number of bonds that are broken on the right and left half of the chain respectively. Thus $\vec{\Delta}$, the deviation of a failure pathway of a heteropolymer from a homopolymer, is given by
\begin{eqnarray}
\vec{\Delta} = \left(
                 \begin{array}{c}
                   n_{L} \\
                   n_{R} \\
                 \end{array}
               \right)
 - \left(
  \begin{array}{c}
    (n_{L} + n_{R})/2 \\
    (n_{L} + n_{R})/2 \\
  \end{array}
\right).\label{Meandering-d-Eqn}
\end{eqnarray}
One can view the dashed line in Fig.\ref{figureX} as a time like co-ordinate. The quantity $\Delta = \vert \vec{\Delta} \vert = \frac{\vert n_{L} - n_{R} \vert}{2}$ is then a random walk as a function of $n_{L} + n_{R}$ along this line. We expect that this quantity scales with the total number of bonds broken $N = n_{L} + n_{R}$ as $\Delta \sim N^{\nu}$, where $\nu$ is the meandering exponent. It is computationally intensive to follow the unzipping histories of long DNA sequences. Moreover, the decision to rupture on the left or right side of the construct is determined by an average over roughly $1/\kappa a$ bonds on each side, thus obscuring the effect of sequence heterogeneity. Nevertheless some insight follows from a simplified bond breaking model which we believe captures the essential physics of the shear unzipping pathways of heterogenous DNA for large $L >> 1/\kappa a$.

\begin{figure}
\includegraphics[width=8cm,height=7cm]{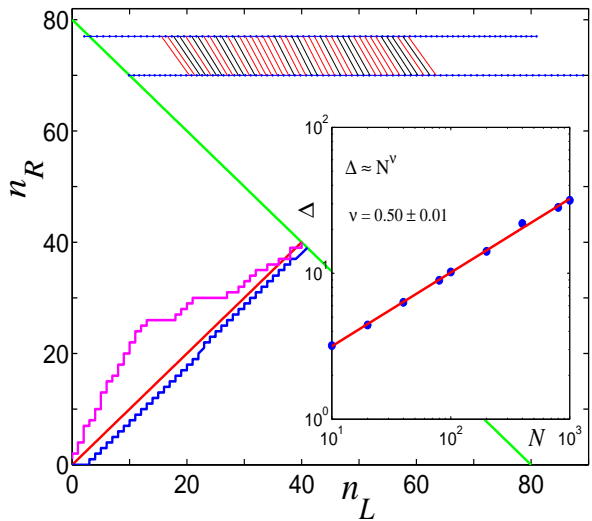}
\caption{\label{UnzippingPathway} Unzipping pathway, parameterized by the number of broken bonds $n_{L}$ on the left vs.\ the number of broken bonds $n_{R}$ on the right half of a randomly chosen heterogeneous DNA sequence with the average ratio of strong (GC) and weak (AT) bonds being 1:1. The GC bond strength is $1.5$ times that of a AT bond (blue line). The system is not large enough to see significant deviations from the pathway of a homopolymer. When the GC bond is made $10$ times stronger, and the bonds representing the backbone $5$ times weaker, larger deviations from the homopolymer unzipping pathway is seen, as indicated by the magenta line. The inset shows the scaling of meandering length (see text) of the failure pathway for a highly simplified model of DNA rupture as a function of system size.}
\end{figure}

The model is constructed from a random DNA sequence that is generated with $1:1$ ratio of $AT$ and $GC$ bonds. Next, the left-most or right-most bond is erased (\textit{i.e.} broken) depending upon which one of these is the weaker one. In case of a tie, one of them is erased randomly. A variant such that both bonds are erased simultaneously gives very similar results. Simulations done on system sizes $L = 10 - 1000$ for both the models yield a value of the meandering exponent $\nu = 0.50 \pm 0.01$, consistent with a one dimensional random walk in the variable $n_{L} - n_{R}$ as a function of the total number of broken bonds $N = n_{R} + n_{L}$ (see Fig.\ref{UnzippingPathway}).

\section{Nonlinear Force extension curves: Semi-microscopic vector model}
\label{forceextensioncurvessemimicroscopicvectormodel}
\begin{figure}
\includegraphics[width=8cm,height=7cm]{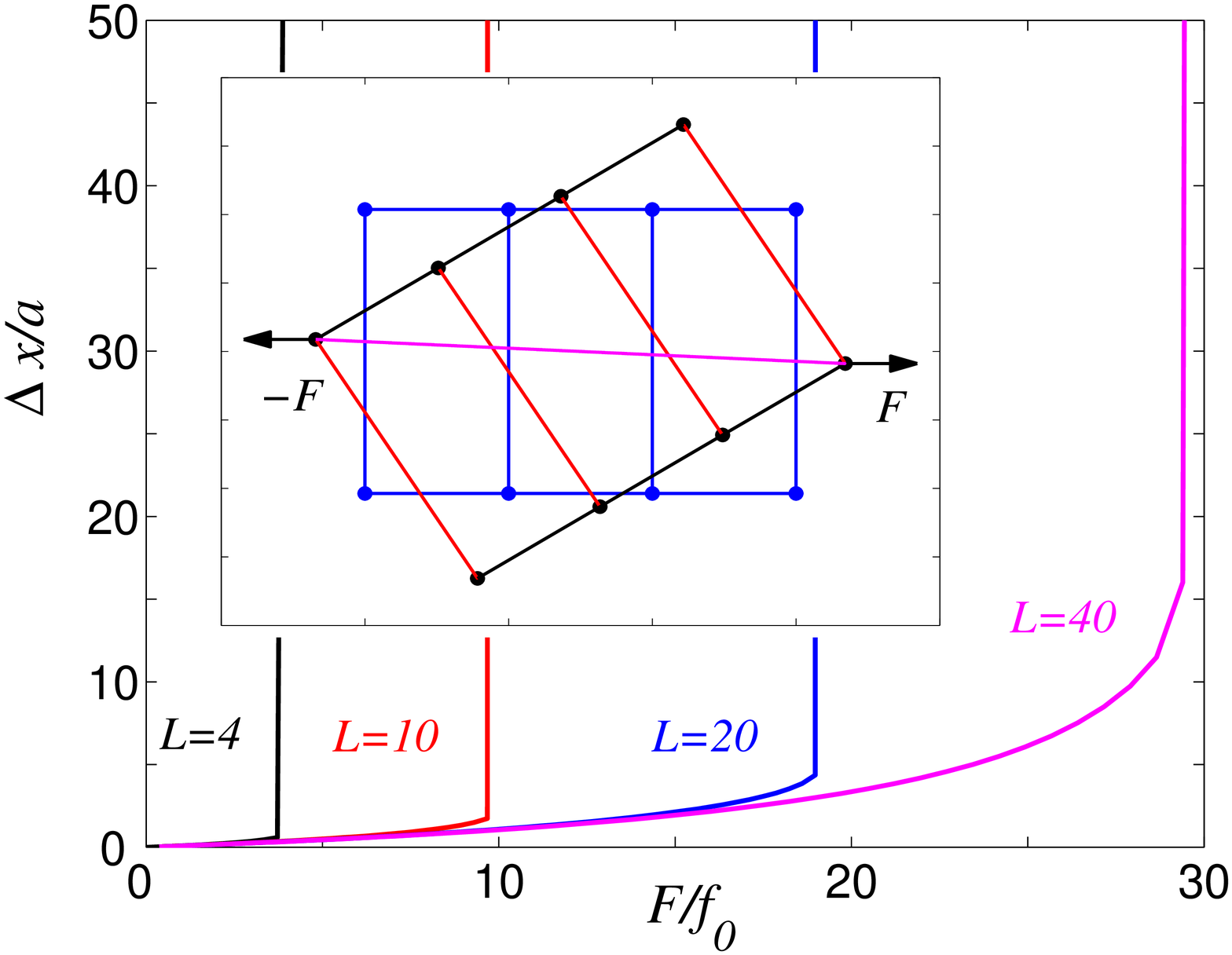}
\caption{\label{FullNonlinearModel} Force vs.\ extension curves for the nonlinear vector generalization of the deGennes model with cross brace for $L=4, 10, 20,$ and $40$ respectively. Inset is a configuration snapshot illustrating the rotation of the sheared DNA for $L=4$ and a small force $F=0.093 f_0$.}
\end{figure}

Figure~\ref{FullNonlinearModel} shows the force versus extension curves arising from the semi-microscopic nonlinear vector model for a homopolymer at zero temperature for a short $L=4$ hybridized DNA segment, for intermediate $L=10$ and $L=20$ DNA hybrids and for a long strand with $L=40$. The force vs.\ extension curves are similar to the ones obtained by the scalar version of the deGennes model. However the rupture force in this case is lower than those obtained in the scalar model without the cross braces but the same $\epsilon$ and $\sigma$ for $V_{LJ}(r)$. A careful examination of the configuration of the chain for a small system ($L=4$) shows that the chain rotates with respect to its initial orientation upon application of collinear forces. In this orientation, the cross braces offer an easier pathway in breaking open the chain. The rupture force obtained from numerics match very well with the estimate of $L$ springs acting in parallel with an effective spring constant of $R^{\prime}_{eff} = 72 \epsilon^{\prime}/(2^{1/3} \sigma^{\prime 2})$ for small systems. Since this rotation leads to softer springs, the deGennes length increases from $\kappa^{-1} = 10 a$ to $\kappa^{-1} = 15 a$ for this system. In our earlier description, we observed that the rupture force saturated for $L \gtrsim 4/\kappa a$. Although we have not simulated large enough systems, we estimate that signatures of the saturation effect appear for $L \gtrsim 60$ in the nonlinear vector generalization of the model.

\section{Concluding Remarks}
\label{concludingremarks}
In conclusion, we have constructed a ``semi-microscopic'' vector description of a sheared double-stranded DNA. The  Hamiltonian Eq.(\ref{Semi-microscopic-H-Eq}), when coarse grained, leads to a nonlinear generalization of a scalar model first studied by deGennes. Cross-brace interactions (designed to represent stacking energies in real DNA) eliminate an unphysical soft mode. Our results show that the deGennes effect, namely the shear strain being localized over a narrow band on either ends of the ladder, arises as well in this more microscopic Hamiltonian. Using this model, we have obtained the force extension curve for a homopolymer at low temperatures. We find a nonlinear force extension curve with an unzipping transition at a critical force that is consistent with roughly $L$ sets of springs (direct interactions and cross braces) acting in parallel for small systems and of order $2/\kappa a$ springs for large systems. We also explored the role of sequence heterogeneity in the scalar version of the deGennes model.

It is of some interest to $(1)$ extend these ideas to the statistical mechanics of finite temperature shear denaturation and $(2)$ study further the effect of sequence heterogeneity. The equations of motion of a sheared chain at finite temperature are given by Eq.(\ref{Equations-Of-Motion-Eq}) with an additive white noise acting on each node of the ladder and obeying the fluctuation dissipation theorem. Thus
\begin{equation}\label{Noise-Statistics-Eq}
\langle \eta_{i}(t) \eta_{j}(t^\prime) \rangle = 2 \Gamma k_B T
\delta_{ij} \delta(t-t^\prime),
\end{equation}
where $i$ and $j$ index particular nodes of the ladder, on both the upper and lower strands. The equations of motion for the $u$-displacement at the $i^{th}$ node is given by
\begin{equation}\label{Equations-Of-Motion-T-Eq}
\frac{\partial u_i}{\partial t} = - \Gamma \frac{\delta H}{\delta
u_i} + \eta_i (t),
\end{equation}
 and similarly for the $v$ field. Preliminary calculations for homopolymers suggest that for this system $\Delta x \sim \frac{k_{B} T}{F - F_c}$, where $F_c$ is the rupture force of the chain, similar to results for unzipping due to forces applied at the same end of a homopolymer DNA duplex,\cite{Lubensky:00,Lubensky:02} provided the chain is much longer than the elastic screening length $\kappa^{-1}$.

Sequence heterogeneity leads to models with elastic ``disorder'', which could dramatically alter the physics of shear
unzipping, as found earlier for tensile unzipping\cite{Lubensky:00,Lubensky:02}. We note that inchworm excitations, which exploit translational invariance,\cite{Neher:04} and are neglected here, are strongly disfavored by sequence heterogeneity. Of particular interest is the shear denaturation pathway described by sequence-dependent bond breaking at alternate ends of a heterogeneous DNA hybrid. It would be interesting to explore in more detail how the deGennes effect is modified for shear denaturation of heteropolymers. Provided the disorder is correlated on the scale of a few base pairs for a double stranded DNA\cite{Buldyrev:95}, we believe that the shear unzipping of a heteropolymer would be similar to unzipping in a tensile mode with sequence heterogeneity. Preliminary results for heterogeneous systems show that the force extension curve comprises of plateaus and jumps akin to tensile unzipping of heteropolymers\cite{Lubensky:00,Lubensky:02}.

We have not incorporated the effects of helical twist in our work here. As discussed above, a more detailed model by Lavery and Lebrun\cite{Lavery:96,Lavery:99} shows that for $3^\prime - 3^\prime$ pulling, DNA unwinds to assume a ladder like form (S-DNA). Recent experiments by Danlowicz \textit{et al.}\cite{Danilowicz:08} have shown that the unzipping force of a DNA hybrid is larger for $3^\prime - 3^\prime$ pulling than that of $5^\prime - 5^\prime$ pulling. A straightforward generalization of the semi-microscopic vector model discussed above that would capture this effect would be to consider two different sets of stacking interaction strengths $\epsilon^{\prime}$ for the diagonals top left corner to bottom right corner as opposed to the ones from bottom left to top right. Since the effective spring constant is that of the cross braces of the leading diagonal, the rupture forces arising from such a model could be different depending upon whether the stronger/weaker cross braces are being broken under shear.

\section{Acknowledgements}
It is a pleasure to acknowledge helpful conversations with C. Danilowicz, E. Kaxiras, M. Fyta, and M. Prentiss. We would especially like to acknowledge the inspiring influence of Pierre-Gilles deGennes. This small elaboration of one of his lesser-known papers can in no way do justice to this giant of twentieth century physics. The authors acknowledge the National Science Foundation, through Grant No. NIRT 0403891, and the National Institutes of Health through Grant No. GM075893 for financial support. BC thanks M. Das and M. Fyta for careful reading of the manuscript. DRN acknowledges financial support from the National Science Foundation, through Grant DMR-0654191, and through the Harvard Materials Research Science and Engineering Center through Grant DMR-0213805. We also acknowledge the Harvard National Nanotechnology Infrastructure Network (NNIN) for computational resources.


\begin{thebibliography}{9999}
\bibitem{footnote} When only nearest neighbor interactions are included and displacements calculated in a Pythagorean fashion the complimentary base pair interactions modeled by a Lennnard-Jones potential is quartic for small displacements. This leads to a compressive force perpendicular to the chain axis which causes the ladder to collapse when sheared. This is akin to a zero shear modulus for a square lattice.
\end{thebibliography}
\end{document}